\documentclass{article}
\usepackage[english]{babel}
\usepackage[margin=0.5in]{geometry}
\usepackage{multicol}
\usepackage{background}

\usepackage{amsmath,amssymb}	
\usepackage{graphicx}			
\usepackage{here}                             

\SetBgScale{4}
\SetBgColor{gray}
\SetBgAngle{90}
\SetBgContents{Draft}
\SetBgPosition{-1.5,-10}
\begin{document}
 
{\centering
 
{\bfseries\Large Plasmon$-$Exciton Interactions Probed Using Spatial Coentrapment of Nanoparticles by Topological Singularities \bigskip}
 
P. J. Ackerman\textsuperscript{1,2}, Haridas Mundoor\textsuperscript{1}, I. I. Smalyukh\textsuperscript{1,2,3,4} and Jao van de Lagemaat\textsuperscript{1,4,5}\\
   {\itshape
\textsuperscript{1}Department of Physics, University of Colorado, Boulder, Colorado 80309, USA \\
\textsuperscript{2}Department of Electrical, Computer and Energy Engineering, University of Colorado, Boulder, Colorado 80309, USA \\

\textsuperscript{3} Liquid Crystal Materials Research Center and Materials Science and Engineering Program, University of Colorado, Boulder, Colorado 80309, USA \\
\textsuperscript{4} Renewable and Sustainable Energy Institute, National Renewable Energy Laboratory and University of Colorado, Boulder, Colorado 80309, USA \\
\textsuperscript{5} National Renewable Energy Laboratory, Golden, Colorado 80401, USA \\

\normalfont (Dated: Nov 14, 2015)
 
   }
}
 
\begin{abstract}

We study plasmon$-$exciton interaction by using topological singularities to spatially confine, selectively deliver, cotrap and optically probe colloidal semiconductor and plasmonic nanoparticles. The interaction is monitored in a single quantum system in the bulk of a liquid crystal medium where nanoparticles are manipulated and nanoconfined far from dielectric interfaces using laser tweezers and topological configurations containing singularities. When quantum dot-in-a-rod particles
\setlength\multicolsep{0pt}
\begin{multicols}{2}

\noindent are spatially colocated with a plasmonic gold nanoburst particle in a topological singularity core, its fluorescence increases because blinking is significantly suppressed and the radiative decay rate increases by nearly an order of magnitude owing to the Purcell effect. We argue that the blinking suppression is the result of the radiative rate change that mitigates Auger recombination and quantum dot ionization, consequently reducing nonradiative recombination. Our work demonstrates that topological singularities are an effective platform for studying and controlling plasmon$-$exciton interactions.
\begin{center}
\includegraphics[width=0.45\textwidth]{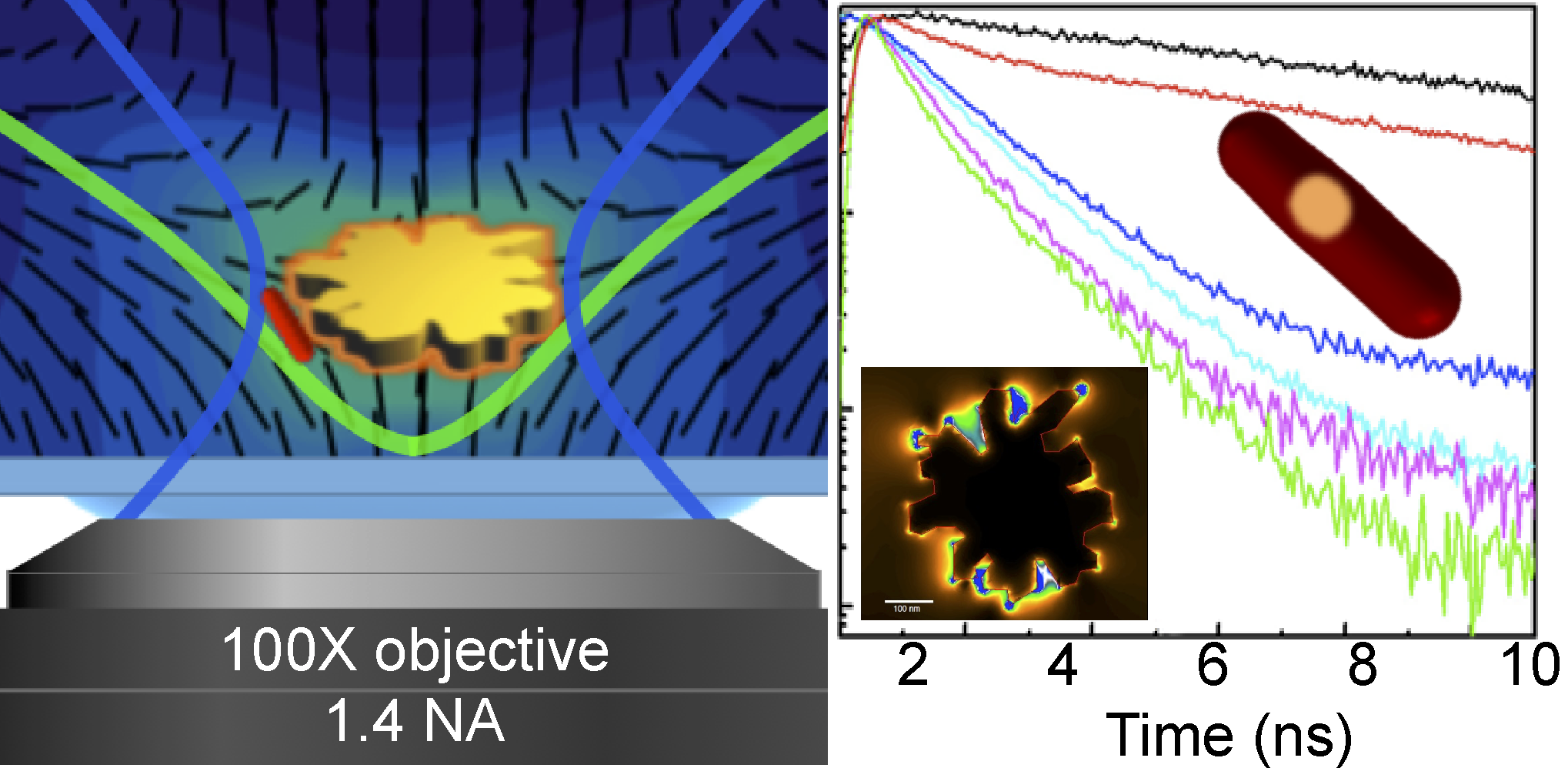} 
\end{center}
\end{multicols}
 
\noindent DOI: 10.1021/acsnano.5b05715
 
\bigskip

\end{abstract}

\begin{multicols}{2}

\section{Introduction}

The possibility of impacting the excited-state dynamics as well as the final fate of excitons by coupling excitons and surface plasmons, possibly leading to desirable outcomes such as increased singlet fission rates or multiple exciton generation, has been an active avenue of research \cite{Johnson2009}. Strong interactions causing Rabi splittings in the 100s of meV are possible between excitons in molecules or quantized nanosized semiconductors and surface plasmons \cite{Gomez2010APL, Gomez2010Nano, Cade2009, Symonds2008}. These interactions can also be used in extremely sensitive single-molecule detection schemes \cite{Zhao2008}. The potential to influence excited-state energies and dynamics is of high interest to exert a measure of control in systems used in solar-to-electricity or solar-to-fuel photoconversion applications. Semiconductor quantum dots are often described as model systems for excitonic behavior. Because of the size quantization, one can shift energy levels in order to, for example, turn-on-and-off singlet fission from molecules that can transfer charges to these nanoparticles \cite{Thompson2014}. Surface plasmons can have strong interaction with excitons on nearby colloidal quantum dots, and this interaction is an interesting approach to manipulating excited state energies \cite{Gomez2010APL, Pelton2015}. One consistent observation when looking at light emission from semiconductor quantum dots is that their fluorescence blinks \cite{Efros2008, Galland2011}. This phenomenon is generally attributed to the dot becoming charged, which makes it dark owing to Auger recombination, and discharging, giving rise to intermittent fluorescence \cite{Efros2008}. A similar effect was observed in current flow through quantum dots trapped in the gap between tip and substrate in a tunneling microscope and can be explained by the same effect \cite{Maturova2013}. It has been shown that it is possible to suppress the fluorescence blinking in quantum dots by controlling the chemical environment \cite{Fomenko2008} and by growing thick, passivating shells around the core of the particles \cite{Wang2009, Mahler2008} giving rise to extraordinarily stable quantum emitters.

In this work, we demonstrate coupling to a nearby plasmonic particle as another way to decrease intermittency which, by increasing the radiative decay rate through the Purcell effect, increases the likelihood of radiative recombination versus ionization of the exciton or Auger recombination and therefore keeps the fluorescent particle in its bright state more often and for longer periods of time. Similar experiments performed on single or assemblies of semiconducting nanocrystals with metal particle or metal surfaces show decrease in the fluorescence blinking \cite{Bharadwaj2011, Fu2007, Ji2015, Karan2015, Ratchford2011, Shimizu2001, Shimizu2002}; however most of the experiments were not performed on the same semiconducting particles throughout the experiments \cite{Fu2007, Ji2015, Karan2015, Shimizu2001, Shimizu2002}. We probe this effect in the bulk of a dielectric medium using a liquid crystal (LC) topological singularity, where we controllably bring a plasmonic particle in close proximity to a single semiconducting nanocrystal using a combination of noncontact optical manipulation and elastic forces, which uniquely allows us to exclude the effects of proximity to substrates on which the plasmon$-$exciton interactions are typically studied \cite{Ratchford2011}. This work also demonstrates that topological singularities

\end{multicols}
\begin{figure}[H]
\begin{center}
\includegraphics[width=0.9\textwidth]{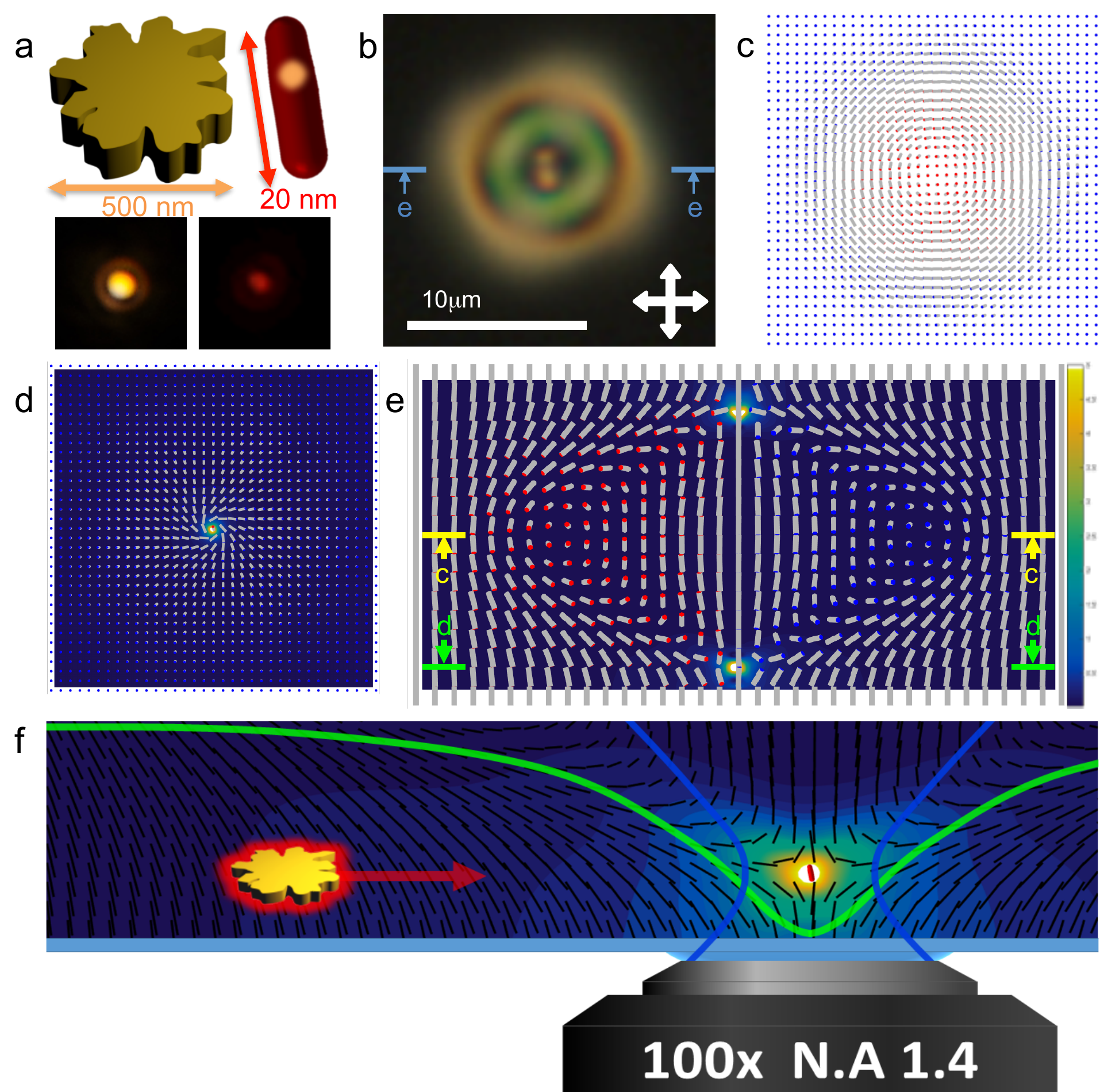}
\end{center}
\caption{Topological singularities as elastic traps for metal and semiconductor nanoparticles. (a) Schematic of a NB (left) and QR (right). The micrographs beneath the schematics show a dark-field microscopy image of a single NB based on the strong scattering from it and a fluorescence image of a single quantum rod; both particles are imaged on a glass substrate. (b) Polarizing optical microscopy texture of a toron; double-headed white arrows depict orientations of polarizers. (c) Computer simulated director structure in the cell midplane, intersecting the toron in its solitonic equatorial plane, as denoted in (e) using yellow lines. (d) Computer simulated director structure within an in-plane cross-section passing through a point singularity of a toron, as denoted in (e), overlaid on the plot of the total elastic free energy density shown using a corresponding color scheme provided in the right-side inset in (e). (e) Computer simulated director structure in the vertical X$-$Z cross-section passing through the toron's central axis and its two hyperbolic point singularities, as denoted in (a) using blue lines. (f) Schematic representation of nanoinclusions interacting with the elastic potential (green curve) formed by a hyperbolic point singularity elastic trap, which is optically probed and observed through an objective lens.}
\label{peFigure1}
\end{figure}
\begin{multicols}{2}

\noindent can serve as an effective platform for studying plasmon$-$exciton interaction and that the fluorescence blinking of semiconducting nanocrystals can be effectively suppressed by using surface plasmons instead of surface passivation.

\section{Results and Discussion}

\subsection{Optical and Elastic Trapping of Nanoparticles}

We use gold Nanoburst \cite{Senyuk2012, Mundoor2014, Senyuk2013} (NB, obtained from Nanopartz Inc.) shaped as $\sim$500 nm discs with irregular sharp edges and dot-in-rod 5$\times$20 nm CdSe/CdS quantum rod (QR) semiconducting nanocrystals with emission peak at 570 nm \cite{Smith2011, Talapin2003}, which are shown schematically and in optical micrographs in Figure \ref{peFigure1}a. Single NBs provide strong and localized field enhancement due to sharp features and are readily identified in dark-field microscopy as strong scattering yellow spots. The QRs that were used have good long-term stability in different chemical environments and can be uniquely identified and imaged based on fluorescence detected using a dichroic mirror and band-pass filter (see Supplementary material for details). These particles are dispersed in a chiral nematic LC host fluid at vanishingly low volume fractions (on the order of 1000 QRs and 10,000 NBs per 1 $\mu$L of the LC) and infiltrated into a LC cell. Spontaneous formation and laser generation of configurations known as torons \cite{Smalyukh2010} (Figure \ref{peFigure1}b) are reliably achieved in regions of cholesteric pitch to cell thickness $p/d \sim 1$ in such cells.

Our optical trapping system can manipulate NBs \cite{Mundoor2014} and self-assembled toron structures with topological singularities \cite{Ackerman2015NatCom}, but not the smaller QRs. This is consistent with the estimates of optical trapping forces as a function of particle dimensions, which show that optical gradient forces for such small QR nanoparticles are comparable in strength to Brownian motion \cite{Ashkin1986}. We therefore control QRs with topological elastic traps of LC singularities that are free of this deficiency and can effectively entrap nanoparticles \cite{Senyuk2012} and molecular self-assemblies \cite{Wang2015}. As illustrated in Figure \ref{peFigure1}, we exploit the hyperbolic point singularities in a frustrated confined chiral nematic LC that are stabilized against annihilation by a solitonic double-twist-torus of a toron shown in a polarizing optical micrograph in Figure \ref{peFigure1}b. Numerical modeling of the director field of the toron structure (Figure \ref{peFigure1}d$-$f) illustrates a large elastic energy density associated with the two twist-bound hyperbolic point singularities, each resulting in a viable elastic trap. Furthermore, these distortions of the director field decay slowly in all lateral directions, so that attractive elastic forces act on particles at distances $\sim$10 $\mu$m much larger than their size \cite{Senyuk2012} and much larger than the range of action of optical trapping forces \cite{Ashkin1986}, thus mediating entrapment of the nanoinclusions that are originally found far from the trapping sites of the point singularities. In the cores of these energetically costly hyperbolic point singularities, the orientational ordering of the LC host is undefined within a volume of effective diameter 10$-$50 nm \cite{Senyuk2012}, which is capable of entrapping even the particles that are too small to be trapped by laser tweezers \cite{Ashkin1986}. By exploiting these properties of hyperbolic point elastic traps, we first localize QRs in the singular cores through elasticity-mediated interactions of the nanoinclusions with the singularities and then observe their fluorescence. Following this, NBs are brought to the vicinity of these point singularities containing QRs using the optical trapping system and then released (Figure \ref{peFigure1}f). Elastic attraction and eventual spatial confinement to the singularity core emerges from a combination of minimization of elastic free energy through sharing of director distortions and displacement of energetically costly LC with reduced orientational order \cite{Senyuk2012}. This results in nanoscale localization and coentrapment of QRs and NBs, allowing us to compare fluorescence of the same QRs when they are entrapped in the singularity alone with the fluorescence of the QRs in close proximity to plasmonic NBs during coentrapment at the latter stage of this experiment.

\begin{figure}[H]
\begin{center}
\includegraphics[width=0.8\textwidth]{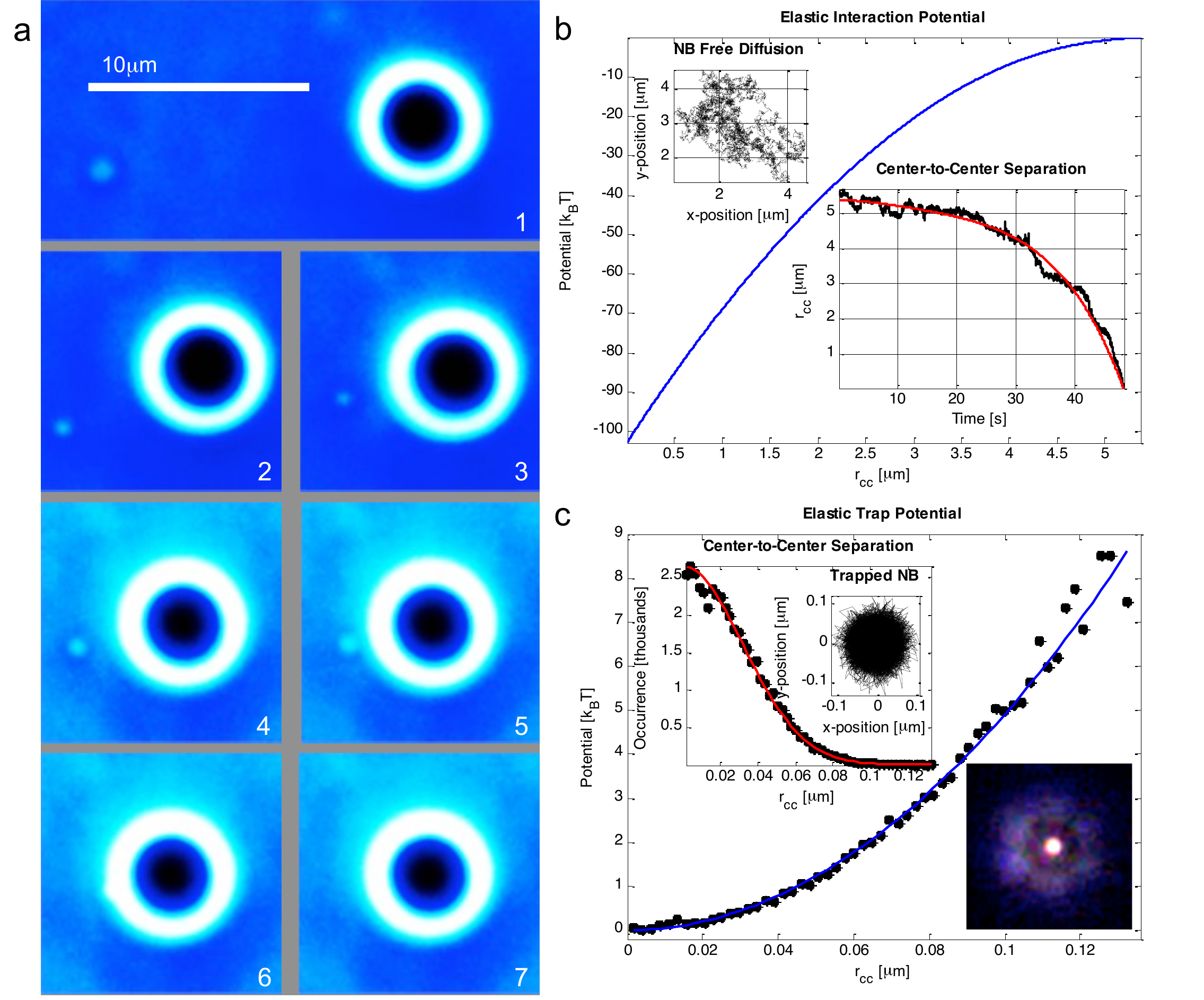}
\end{center}
\caption{Release from optical traps, elastic interaction, and entrapment of nanoparticles in a topological point singularity. (a) Optical and elastic trapping of a NB (small bright spot) nearby a toron (visible as large bright ring of enhanced light scattering). Frames 1$-$3 show optical manipulation of the NB to a desired initial location near the toron and close to one of the substrates substrate. Frames 4$-$7 show the NB elastically pulled toward the toron center after its release from the laser trap. (b) Elastic interaction potential versus the center-to-center separation ($r_{cc}$) between the NB and toron singularity (blue curve), derived from the bottom-right inset plots of $r_{cc}$ versus time and using the NB diffusion constant obtained from probing free diffusion (top-left inset). (c) Trapping potential and stiffness characterized via tracking the singularity-entrapped NB. The trapping potential is derived from the $r_{cc}$ histogram shown in the top-left inset. The bottom-right inset image shows the NB particle entrapped by a toron singularity as observed in dark-field mode (the inset edge length is $\approx$5 $\mu$m). The characteristic stiffness of a toron singularity elastic trap is $\sim$4$\times10^{-3}$ pN/nm.}
\label{peFigure2}
\end{figure}

To characterize topological elastic trapping of nanoparticles, the NB-singularity interaction was probed with video microscopy (Figure \ref{peFigure2}). Frames 1$-$3 of Figure \ref{peFigure2}a show the NB manipulated by our optical trapping system. Frames 4$-$7 of Figure \ref{peFigure2}a reveal the attractive NB-singularity elastic interaction after the NB is released near one of the point singularities of a toron. Using dark-field video microscopy, we optically track the NB far from singularities (free diffusion), as it is pulled into a topological elastic trap and also once it is trapped in a singularity. We derive the elastic interaction potential and elastic trap potential. The erratic 2D free diffusion trajectory (Figure \ref{peFigure2}b) in a uniformly vertically aligned LC yields the viscous drag coefficient $c_d$. Since this system is in the regime of low Reynolds number (Re $\ll$ $10^{-7}$) \cite{Senyuk2012}, inertia effects are negligible and the NB-singularity elastic attractive force $F_{el}$ is balanced by the viscous drag force, $F_d = c_dv(t)$, where $v(t) = dr_{cc}/dt$ and $r_{cc}$ is the NB-singularity time dependent center-to-center separation (Figure \ref{peFigure2}b). The elastic interaction potential is then calculated by integrating $F_{el}$ obtained from this force balance (Figure \ref{peFigure2}b). By tracking the NB trapped by the singularity (Figure \ref{peFigure2}c), we construct a histogram of NB displacements, from which we derive the elastic trap potential by assuming a simple harmonic potential attraction model near the equilibrium location \cite{Ackerman2015NatCom}. This reveals that the center of the NB is confined within a volume of effective radius smaller than 100 nm.

\subsection{Plasmon$-$Exciton Interactions} 

\begin{figure}[H]
\begin{center}
\includegraphics[width=0.65\textwidth]{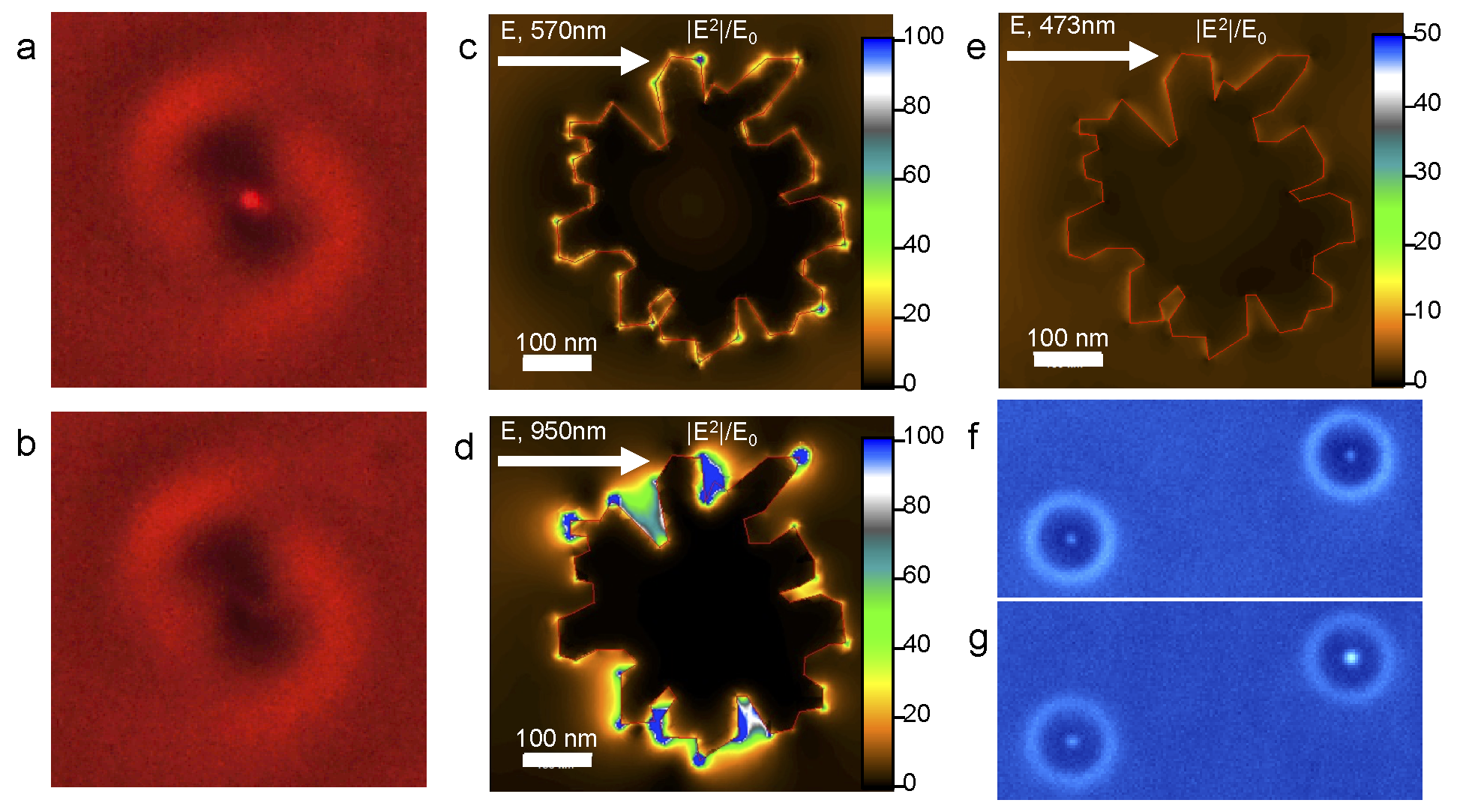}
\end{center}
\caption{Fluorescence enhancement and blinking of QRs without and with NB coentrapment. (a,b) Bright field micrographs depicting the fluorescence signal of a QR elastically trapped in a toron point singularity while in the ``ON'' (a) and ``OFF'' (b) states. (c$-$e) Computer-simulated electromagnetic field enhancement around NBs calculated at 570, 950 and 473 nm excitation wavelengths of the incident light with linear polarization (vibration direction of the light's electric field E) by a NB in the LC medium. (f,g) Time averaged image of QRs elastically trapped in torons before (f) and after (g) introducing a NB into the toron on the right.}
\label{peFigure3}
\end{figure}

\begin{figure}[H]
\begin{center}
\includegraphics[width=0.65\textwidth]{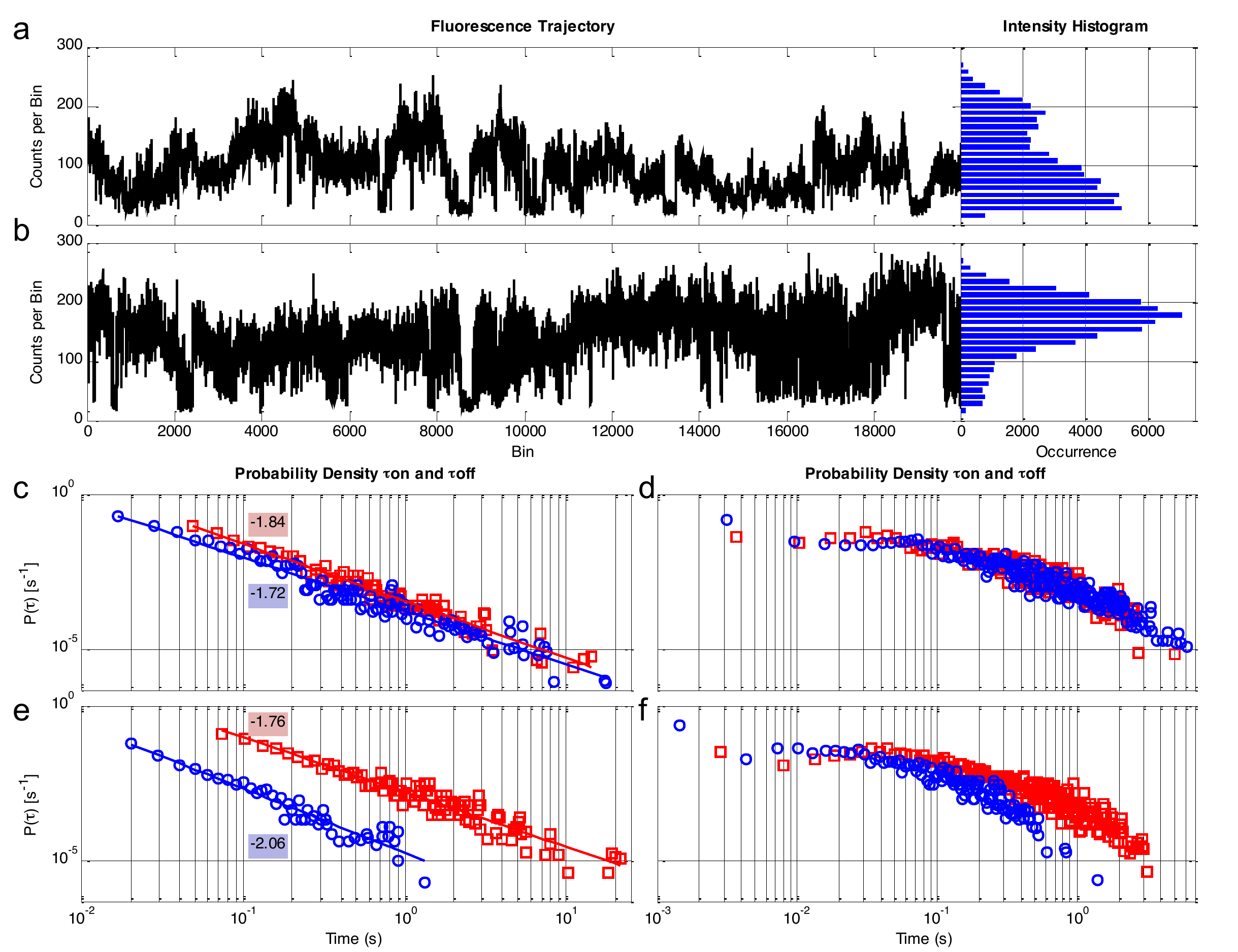}
\end{center}
\caption{Characterization of fluorescence intermittency of QRs before and after introducing plasmonic NBs. (a,b) Fluorescence time traces of QRs elastically trapped in a toron point singularity without (a) and with (b) the NB particle in the elastic trap. The ``ON'' and ``OFF'' times of the QRs is represented by the corresponding histograms in the right-side insets. Analysis of fluorescence time trace with constant thresholding (c,d) and using the change point analysis method (e,f) for the curves displayed in (a) and (b) and corresponding to (c,e) the QRs alone and (d,f) in the close neighborhood of NBs. Here $\rho(\tau)$ is probability density defined as in reference \cite{Kuno2000} and plotted as $\rho(\tau$ON) and $\rho(\tau$OFF) for on and off times (red squares and blue circles, respectively). $\tau$ON and $\tau$OFF are time periods of sustained fluorescence emission and darkness, respectively.}
\label{peFigure4}
\end{figure}

To probe the photo physics of QRs, torons containing only QRs in only one of their singularities are illuminated with a mercury lamp and the resulting fluorescence is monitored. Figure \ref{peFigure3}a and \ref{peFigure3}b show bright field micrographs of such a toron with a QR in the ``ON'' and ``OFF'' state, respectively. Then, to probe the effects of plasmon$-$exciton interaction on these QR properties, the NB is cotrapped by the point singularity according to the procedure described above. Computer simulations (Figure \ref{peFigure3}c$-$e) reveal the electromagnetic field enhancement by the NB at wavelengths 570 nm, 950 and 473 nm, which are the emission and two excitation wavelengths used in the experiments discussed below. Consistently, the QR emission increases (Figure \ref{peFigure3}f,g), as seen from comparing time-averaged frames of two torons with QR particles before and after coentrapping the NB in the point singularity (top right). To analyze the corresponding blinking characteristics of QRs, we record fluorescence trajectories using 473 nm diode laser illumination and Avalanche photo diode (APD) with a binning time of 10 ms. Figure \ref{peFigure4}a,b show characteristic fluorescence trajectories of QRs before and after introducing a NB particle into the elastic trap. It is clear that QR particles stay predominantly in the ``ON'' state after cotrapping NBs in the LC singularity. It should be noted that the blinking trajectories are affected by the diffusion of the particles, resulting in slight intensity fluctuations due to displacement of the particles from the focal plane. The effects of introducing the NB are clearly visible from the histogram plotted in the right-side insets of Figure \ref{peFigure4}a,b, showing a change in the blinking statistics. Further, we have performed detailed analysis of the blinking following the constant thresholding \cite{Shimizu2001, Crouch2010, Kuno2000}

\end{multicols}
\begin{multicols}{2}

\noindent and change point analysis methods reported recently \cite{Schmidt2014}, where we have calculated the probability density for the ``ON'' and ``OFF'' times (Figure \ref{peFigure4}c$-$f). The results of our analysis using the former method are consistent with the earlier reports \cite{Shimizu2001, Crouch2010, Kuno2000} showing typical power-law behavior of such data (Figure \ref{peFigure4}c,e) and that the particle is far more often in the ``ON'' state in the presence of plasmonic enhancement than without it. The same is concluded from the change-point analysis (Figure \ref{peFigure4}d,f). The probability density of ``ON'' states of QRs in the presence of a NB increases, indicating decreased blinking and increased prevalence of the ``ON'' state.

\begin{figure}[H]
\includegraphics[width=0.49\textwidth]{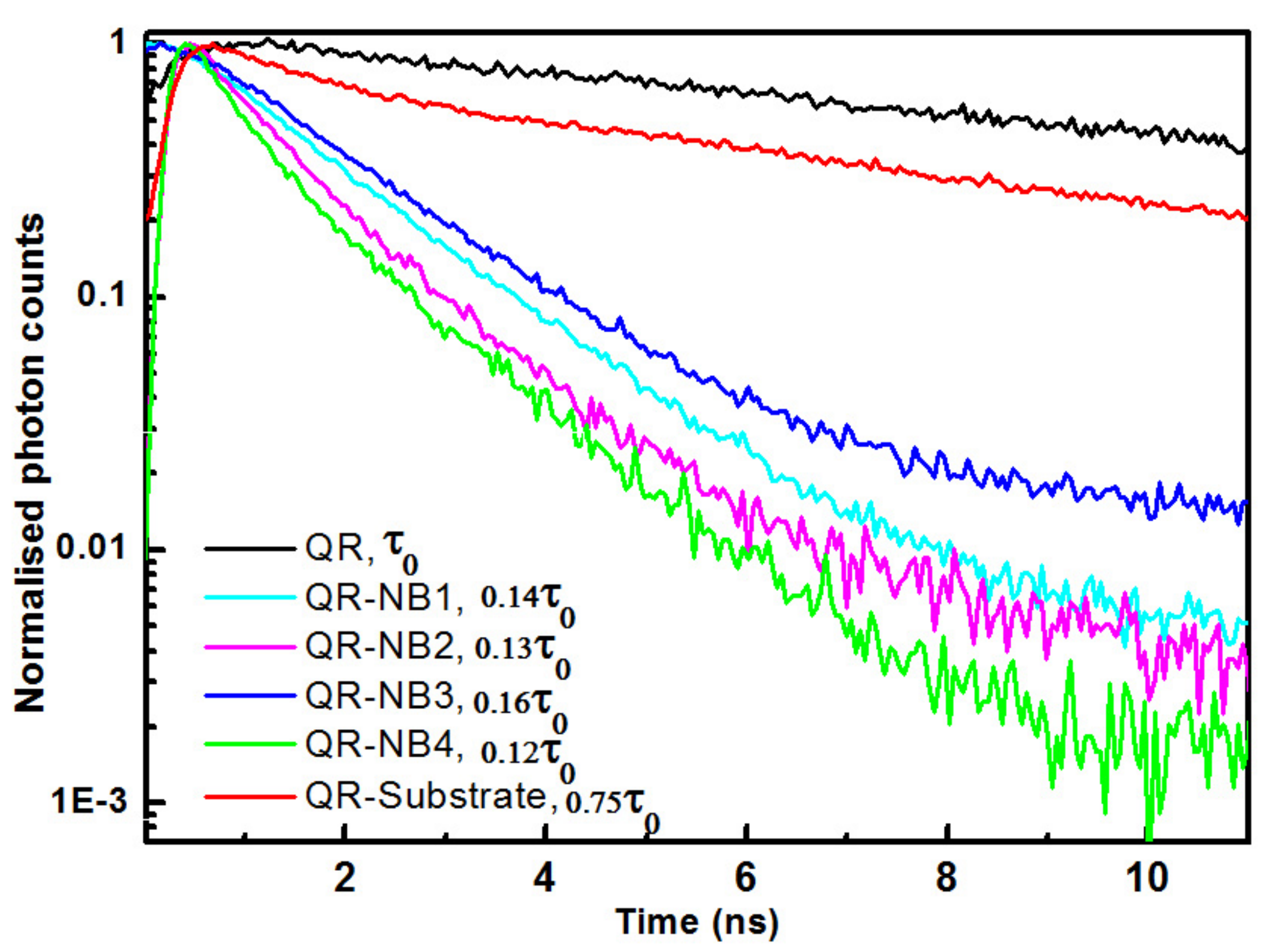}
\caption{Characterization of fluorescence lifetime of QRs. Photoluminescence decay for QRs entrapped in a toron's point singularity (black curve) and when placed on a glass substrate (red curve) are compared to four sets of characteristic decay curves for QRs in topological point singularities coentrapped with the NB (blue, cyan, magenta and green curves). The fluorescence lifetime for QRs measured in the topological elastic trap is $\tau_0 \approx 10.16$ ns.}
\label{peFigure5}
\end{figure}

To gain additional insights to the nature of plasmon$-$exciton interactions, we have performed fluorescence lifetime measurements of QRs located inside a point singularity, before and after introducing the NBs. For this, the excitation of the system was realized through a two photon absorption process using 950 nm light from a pulsed Ti:sapphire laser. Although the accuracy of lifetime measurements is somewhat limited by the repetition rate of our laser (80 MHz), the comparison of the decay curves unambiguously shows that the photoluminescence from the QR decays much faster in the presence of NBs than when QR are entrapped alone or on a dielectric substrate (Figure \ref{peFigure5}). The excited-state lifetime for the QR isolated through trapping within the topological singularity in the bulk of the dielectric LC medium is $\approx$10.16 ns. Introduction of a NB into the trap for different experiments with different nanoparticles repeatedly results in a dramatic change of this lifetime by a factor ranging from 0.12 to 0.16, yielding lifetimes within 1.2$-$1.6 ns. The somewhat broad range of lifetime values observed for different QR-NB combinations is likely due to the irregular shapes of the NBs and different specific locations/orientations of the QRs relative to the NBs, as well as variations in the illumination light's propagation direction relative to the QR-NB system. The blinking experiments were consistently performed on one and the same QR

\end{multicols}
\begin{multicols}{2}

\end{multicols}
\begin{multicols}{2}

\noindent before and after introducing a NB, the fluorescence lifetimes were measured for a series of QR-NB assemblies produced in the toron's point defects in each case comparing the same QR before and after addition of a NB particle. Importantly, the observed dramatic effect of plasmon$-$excition interaction on the fluorescence lifetime is separated from the effect of surface proximity to a dielectric substrate such as glass. Interaction with the glass substrate results in a much smaller decrease of the fluorescence lifetime, $\approx25\%$ (Figure \ref{peFigure5}).

Our experimental finding of nearly an order of magnitude increase in the fluorescence rate of QRs upon adding the plasmonic NB to an elastic trap is consistent with the Purcell effect, owing to the increased density of final states for photon emission in the presence of the NB \cite{Song2005, Anger2006, Govorov2006, Kuhn2006}. Such an effect also explains the decrease in blinking. The intensity of the emission for the ``ON'' state is not strongly affected and since the QR particle is 10$-$20 nm from the NB surface, where nonradiative rates are only marginally affected \cite{Anger2006}, one does not expect to observe any significant quenching of the fluorescence. Second, the field enhancement produced by a NB at the excitation wavelength used for blinking experiments (Figure \ref{peFigure3}e) is small and has minimal effect on the excitation rate of QRs. The coupling of the QR emission to the plasmon modes of the NBs influences the fluorescence blinking by reducing the probability of the dark exciton generation and by driving the branching of the excited states of the QR to light emission instead of charge carrier trapping. This keeps the QRs predominantly in the ``ON'' state instead of the ionized ``OFF'' state. Similar but usually smaller enhancements of the fluorescence decay and suppression of blinking have been observed before in chemically pre-engineered nanoparticles with plasmonic resonator shells around the quantum dot \cite{Ji2015, Karan2015}, quantum dots immobilized on plasmonic substrates \cite{Fu2007, Wu2010} and in engineered quantum dot/plasmonic particle dimers \cite{Cohen-Hoshen2012}. In these experiments, however, the effect of the plasmonic field cannot be probed using the same quantum dots at different experimental stages, making it difficult to unequivocally prove that the effect is due to the plasmonic particle's presence. In contrast, in a similar experiment to ours \cite{Ratchford2011}, a plasmonic particle was brought closer to a colloidal core/shell quantum dot on a glass substrate using an AFM tip and a reversible suppression of blinking and a decrease of the fluorescence lifetime comparable to that described here was observed, albeit the surface proximity could have additionally influenced this behavior and in a different study \cite{Bharadwaj2011}, a strong modification of nonradiative rates leading to quenching of the photoluminescence and consequent suppression of blinking was observed when a gold particle suspended on a tip was brought into closer proximity than used in this work to a quantum dot. In that case the suppression of blinking was due to an increase in nonradiative rates which also would lead to a decrease in ionization events. In the current work, we show that this effect can be seen even for the same plasmonic-quantum nanoparticle system suspended in its entirety in a fluid medium bulk, where the plasmonic field was specifically brought to the quantum nanoparticle by ``soft'' self-assembly methods instead of external forces. While the sharp edges of the NB help to localize the electromagnetic field very strongly and help make the NB easy to move using laser tweezers, the randomness of the shape of the NB limits our knowledge of the QR position and orientation relative to NB enhancement field, resulting in variations in the plasmon$-$exciton interaction for each individual NB-QR combination and somewhat limiting the possibility of further quantifying the effect, which will be the goal of our future studies.
 
Interestingly, with the exception of rare cases, coentrapment of QRs and NBs in the topological point singularities is such that the fluorescence of QRs is enhanced, the lifetime decreases, and the ``ON'' states are more prevalent. This may imply that the QRs are entrapped in the region of elastic distortions around the somewhat bigger NB rather than physically touching it, in which case one would expect quenching of fluorescence. Although chemical methods allow for producing nanostructures with well-defined distance between plasmonic metal and semiconductor components pre-engineered to provide enhancement \cite{Liu2006, Khanal2012}, it is interesting that the NB-QR nanostructures achieve this through elasticity-mediated self-assembly driven by the LC topological singularity. This, along with the recent progress in generating stable arrays of topological singularities \cite{Evans2013, Ackerman2012}, opens opportunities of mass-producing LC-based nanostructured materials with arrays of singularities and self-assembled metal$-$semiconductor nanostructures with pre-engineered plasmon$-$exciton interactions and the ensuing new physical properties. We also note that the conditions for plasmon$-$exciton interactions studied here have not been optimized for particular outcomes of potential interest, such as maximization of fluorescence ``ON'' time or shortening of photoluminescence lifetime. However, the comparison of computer-simulated plasmonic field enhancement at different wavelengths (Figure \ref{peFigure3}), combined with preselected spectral characteristics of quantum rods or dots, shows ample possibilities for engineering the physical behavior of such nanostructured self-assembled systems, which we will pursue in future studies.

\section{Conclusions}

We have demonstrated that topological singularities can be used to coentrap quantum nano-rods and plasmonic nanoparticles within a highly localized volume of a dielectric LC medium, overcoming limitations of laser manipulation techniques that lack the ability to trap quantum dots. This allows us to study plasmon$-$exciton interactions within the bulk of the LC fluid, unaffected by the interaction with substrates. Moreover, trapping of the nanoparticles in LC topological singularities allows us to probe effects of plasmon$-$exciton interactions through characterizing fluorescence of the same semiconducting particle before and after introducing the plasmonic particle. Characterization of emission properties of studied quantum rods before and after coentrapment with plasmonic nanoparticles indicates an order of magnitude decrease in the fluorescence lifetime, consistent with the Purcell effect, enhanced brightness and an increased persistence of the ``ON'' state. Our findings not only enhance the current understanding of plasmon$-$exciton interactions but also point to the possibility for exploiting such interactions in designed self-assembled mesostructured composite materials with pre-engineered physical behavior. Furthermore, considering the facile response of liquid crystals to external fields and light \cite{Liu2014}, such interactions and the ensuing composites can be dynamic and reconfigurable.

\section{Materials and Methods}

\subsection{Chiral Nematic LC Preparation}

We mix 4-cyano-4'-pentylbiphenyl (5CB) nematic host (Chengzhi Younghua Display Material Co., Ltd.) with a chiral agent, either CB15 (EM Industries) or cholesteryl pelargonate (Sigma-Aldrich). The desired cholesteric pitch p, the distance over which the LC ground state director twists by 2$\pi$, is defined by controlling the composition of the LC mixture according to the relation $p = 1/(|H_{HTP}|\cdot C_{agent})$, where $H_{HTP}$ is the helical twisting power in 5CB (7.3 $\mu m^{-1}$ for right-handed CB15 and $-$6.25 $\mu m^{-1}$ for left-handed cholesteryl pelargonate where the $\pm$ sign denotes the corresponding right/left handedness) and $C_{agent}$ is the concentration of the chiral additive in the nematic host. All experiments are performed using LCs with $p \approx 8$ $\mu$m and the handedness was found to play no role in the following experiments. By mixing the LC with either NBs dispersed in ethanol or QRs in toluene and then slowly evaporating the solvent at 65 $^\circ$C, we prepare two separate NB-LC and QR-LC colloidal nanoparticle dispersions, as needed for the experiments described above.

\subsection{LC Cell Fabrication}

LC cells were assembled using glass plates treated with Dimethyloctadecyl[3-(trimethoxysilyl) propyl] ammonium chloride (Sigma-Aldrich) to obtain strong perpendicular surface boundary conditions for the LC director describing the local average orientation of rod-like molecules. Wedge cells with thickness $d$ = 7$-$10 $\mu$m were produced by spacing two substrates with glass fiber segments of diameter 7 and 10 $\mu$m in UV-curable glue at the opposite ends of the cell. The dihedral angle between the substrates was kept below 3 degrees such that the local thickness variation can be neglected. The LC mixtures were coinfused into the cell by capillary forces at 75 $^\circ$C, rapidly cooled to room temperature and sealed with fast-setting epoxy.

\subsection{Experimental Setup and Procedures}

The experimental setup was built around an inverted microscope IX81 (Olympus). The optical trapping system, used for noncontact optical manipulation of NBs and torons, is based on a diode laser (1,064 nm, from LaserGlow) \cite{Senyuk2012, Mundoor2014}. Excitation of QRs is done using a mercury lamp, a 473 nm diode laser (from LaserGlow) or pulsed excitation from a tunable Ti:sapphire oscillator (Chameleon Ultra II, Coherent, 140 fs, 80 MHz, used at 950 nm) and the fluorescence signal is detected using an Avalanche photo diode (APD, Picoquant). Fluorescence is detected through a confocal pinhole and filtered by a band-pass filter (centered at 610 nm, 75 nm bandwidth, from Chroma Technology Corp.). Light is focused by an aspheric lens to the active area of the APD connected to a data acquisition board (NIDAQ-6363, National Instruments) and photon counting hardware (SPC-130, Becker $\&$ Hickl GmbH). We use a 100$\times$ oil immersion objective with adjustable numerical aperture (NA) of 0.6$-$1.3, an electron multiplying charge coupled device (EMCCD, iXon3 888, Andor Technology) and a CCD camera (Flea, PointGrey) for imaging. Dark-field imaging is achieved by adjusting the NA of the 100$\times$ objective to be smaller than the NA of a dark-field condenser, allowing us to visualize strong scattering of individual NBs, which we use for tracking of these nanoparticles. The fluorescence blinking of QRs is analyzed based on the thresholding method and also using a change-point algorithm \cite{Watkins2005}, as discussed in more detail in Supplementary material.

\subsection{Numerical Modeling}

Equilibrium LC director configurations are modeled through minimization of elastic free energy \cite{Ackerman2015NatCom} (Figure \ref{peFigure1}c$-$e) assuming strong perpendicular surface boundary conditions (see Supplementary material for details). Elastic constants of 5CB and experimental geometric parameters such as $d/p = 0.9$ are used in this modeling. The electromagnetic field enhancement of NBs in LCs is simulated using commercially available software (COMSOL Multiphysics), which is based on the finite element method, with a detailed description of the simulation approach provided in Reference \cite{Mundoor2014}. Briefly, the incident electromagnetic wave passes through the computational volume from above (Z direction) along the orientation of the LC director. The simulation is done over a cylindrical volume with the lateral diameter of 1.5 $\mu$m and height of 1 $\mu$m, with the NB located at the center, enclosed between perfectly matched layers in the radial directions and optical ``ports'' in the vertical direction of the cylindrical volume. The simulations were performed at the two excitation wavelengths as well as the emission wavelength of the QRs.

\end{multicols}

\section{Acknowledgments}
~~~ We thank C. Schoen and Nanopartz Inc. for providing NBs as well as J. M. Luther for providing QRs used in studies presented in this work. We acknowledge technical assistance of T. Lee and B. Senyuk. I.I.S. is grateful for the hospitality of NREL during his sabbatical stay. We acknowledge support of the Division of Chemical Sciences, Geosciences, and Biosciences, Office of Basic Energy Sciences of the US Depart- ment of Energy under Contract No. DE-AC36-08GO28308 with the National Renewable Energy Laboratory (P.J.A., J.v.d.L and I.I.S. during his sabbatical leave) and also partial support of the NSF grant DMR-1410735 (H.M. and I.I.S.).
Supporting

\section{Plasmon$-$Exciton Interactions Probed Using Spatial Coentrapment of Nanoparticles by Topological Singularities:  Supplementary material}	
\label{plasmonexitonSI}

\begin{enumerate}

\item
Change point analysis.

We use a change point detection algorithm to locate potential fluorescence intensity change points in the fluorescence trajectories obtained for the blinking QRs \cite{Schmidt2014}. This is a rigorous and quantitative method based on a standardized weighted generalized likelihood ratio test for detection of discrete intensity jumps, known as change points, and binary recursion to locate all potential change points \cite{Watkins2005}. The realization of this method was implemented using home-made Matlab-based software. Detected photons will be Poisson distributed and the probability that there is a change point in the data can be compared to the probability there is not a change point. A confidence level and region for all of the possible change points can be assigned. Once change points for a given confidence level are located, intensity values are assigned to regions of the trajectory without intensity changes. Data collected from NIDAQ hardware at a sample rate of 2M samples per second were recorded. The raw (un-binned) data were either integrated over time to generate bin values for the threshold method or used in raw form for statistical change point detection analysis. The main benefit of the change point detection is to remove any artifacts of time resolution artificially imposed by binning the raw data. 

\item
Details of experimental setup and procedures.

Optical manipulation and video microscopy of nanoparticles and torons and the experimental studies of photoluminescence were performed with setups built around inverted microscope IX81 (Olympus). An optical trapping system based on a diode laser source operating at a wavelength of 1064 nm (from LaserGlow) and integrated with laser illumination (diode laser source operating at a wavelength of 473 nm from LaserGlow). Fluorescence signal was detected with an Avalanche photo diode (APD, Picoquant). A fluorescence cube was used with a band-pass ``clean up'' filter (notch center/width 480/20 nm) for laser input, a dichroic mirror (long pass 495) for separating the illumination from fluorescence signal and further filtered by a band-pass filter (notch center/width 610/75 nm) along the detection path. A large (100 $\mu$m) confocal pinhole and aspheric lens were used to select and focus fluorescence signal to the active area of the APD. Detected photons were counted using a data acquisition board (NIDAQ-6363, National Instruments) and homemade Matlab software. A pulsed excitation source from a Ti:Sapphire oscillator (Chameleon Ultra II, Coherent, 140 fs, 80 MHz tuned to 950 nm) and detection of fluorescence through a side imaging port was limited by a confocal pinhole, filtered by a band-pass filter (notch center/width 610/75) and focused by an aspheric lens to the active area of an APD connected to photon counting hardware (SPC-130, Becker $\&$ Hickl GmbH) for lifetime measurements. We utilized a 100x oil immersion objective with adjustable numerical aperture (NA) of 0.6$-$1.3 with an electron multiplying charge coupled device (EMCCD, iXon3 888, Andor Technology) or a CCD camera (Flea, PointGrey) at a rate of 60 fps for imaging. 

\item
Details of numerical modeling of equilibrium director structures.

We use numerical minimization of free energy to obtain the equilibrium configurations of the director field $\bf{N}(\bf{r})$ in confined chiral nematic LCs. The 3D toron structures are modeled based on the minimization of bulk Frank-Oseen elastic free energy $F$ assuming infinitely strong vertical surface boundary conditions and initial conditions of a loop of double twist.

\begin{multline}
F_{elastic} = \frac{K_{11}}{2}(\nabla \cdot \mathbf n)^2+\frac{K_{22}}{2}(\mathbf n \cdot \nabla \times \mathbf n \pm \frac{2\pi}{p})^2+\frac{K_{33}}{2}(\mathbf n \times \nabla \times \mathbf n)^2-K_{24}\nabla \cdot (\mathbf n (\nabla \cdot \mathbf n) + \mathbf n \times \nabla \times \mathbf n)
\label{eq:fePESI}
\end{multline}

Where $K_{11}$, $K_{22}$, $K_{33}$, and $K_{24}$ are elastic constants describing splay, twist, bend and saddle splay deformations, respectively. The saddle-splay constant $K_{24}$ is difficult to measure experimentally \cite{Polak1994, Anderson1999}. We assume $K_{22}=K_{24}$, in agreement with previous studies \cite{Polak1994, Anderson1999, Smalyukh2010}. Through numerical modeling, we find that torons in cells with thickness $d$ in micrometer range are ground-state structures for $d/p \approx 1$ even when this term is not taken into account (at $K_{24}=0$), although taking $K_{22}=K_{24}$ further assists to stabilize these skyrmionic field configurations with point defects. We therefore set $K_{22}=K_{24}$ while using all other experimentally measured constants of the used liquid crystal (Table 1). 
To find the equilibrium director field, minimization of the free energy is computed with a relaxation method \cite{Anderson1999}. The equilibrium state of $\bf{N}(\bf{r})$ have functional derivatives of the free energy $\delta F/\delta n_i$, where $n_i$ is the projection of the director $\bf{N}(\bf{r})$ onto the $i$-axis ($i$= 1(x), 2(y), 3(z)). From a numerical point of view, the spatial derivatives of $\bf{N}(\bf{r})$ are computed using the $2^{nd}$ order finite difference scheme in a volume broken up into a rectangular computational grid. Periodic boundary conditions are applied along the $\hat x$- and $\hat y$-directions while fixed vertical boundary condition are used along the $\hat z$-direction. At each time step $\delta t$, $\delta F/\delta n_i$ and the resulting elementary displacement $\delta n_i$ defined as $\delta n_i=-\Delta t\frac{\delta F}{\delta n_i}$ are computed. The maximum stable time step used in the relaxation routine is determined as $\Delta t=\frac{min(h_i)}{2max(K)}$, where $min(h_i)$ is the smallest computational grid spacing and $max(K)$ is the largest elastic constant. Steady state is determined through monitoring the change with respect to time of the spatially averaged functional derivative. When this value asymptotically approaches zero, the system is assumed to be in equilibrium. The discretization is done on a large grid $(103 \times 103 \times 52)$, which is important to assure that the minimum-energy $\bf{N}(\bf{r})$ is indeed a structure localized in space in equilibrium with the surrounding untwisted LC and that the periodic boundary conditions do not introduce artifacts affecting stability. Using grid spacing of $h_x = h_y = h_z = 0.15$ $\mu$m and 52 grid points across the cell gives sample thickness $d = 7.80$ $\mu$m, comparable to that used in experiments. All presented simulations have been computed for $d/p = 0.9$.
The free energy value was calculated using eq. \ref{eq:fePESI} and the second order centered finite difference scheme for each point within the volume slice. Plotting the free energy density of slices of 3D simulation data was achieved using Matlab's built in contour function, obtaining the elastic free energy landscapes such as the one shown in Fig. \ref{peFigure1}e for the vertical cross-section and in the supporting figure \ref{peFigure1SI} for the lateral cross-section containing the hyperbolic point defect.

\end{enumerate}

\begin{center} 
Table 1:  Material parameters of the nematic 5CB host used in computer simulations. \\
\begin{tabular}{lcccc}
\hline \hline
LC$\backslash$Properties & $K_{11},pN$&$K_{22},pN$&$K_{33},pN$&$K_{24},pN$ \\ \hline
5CB              & ~6.4       & 3.0~        & 10.0        & 3.0~      \\
\hline \hline
\end{tabular}
\end{center}

\begin{figure}[H]
\begin{center}
\includegraphics[width=0.3\textwidth]{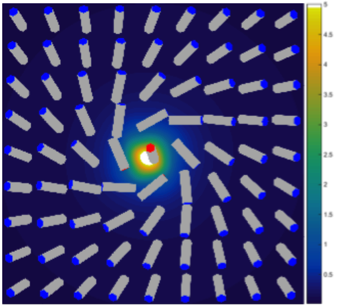}
\end{center}
\caption{Computer simulated director structure within an in-plane cross-section zoomed into and passing through a point singularity of a toron, overlaid on the plot of the total elastic free energy density shown using a corresponding color scheme provided in the right-side inset.}
\label{peFigure1SI}
\end{figure}

\end{document}